\documentclass[english,prd,twocolumn,showkeys,superscriptaddress]{revtex4}
\usepackage[latin1]{inputenc} 
\usepackage{fontenc}      
\usepackage{graphicx,epsfig,units}

\newcommand{\bean}{\begin{eqnarray}}
\newcommand{\eean}{\end{eqnarray}}
\newcommand{\be}{\begin{equation}}
\newcommand{\ee}{\end{equation}}

\begin{document}

\title{An inhomogeneous alternative to dark energy?}
\date{\today}

\author{H\aa{}vard Alnes}
\email{havard.alnes@fys.uio.no}
\affiliation{Department of Physics, University of Oslo, PO Box 1048
  Blindern, 0316 Oslo, Norway} 

\author{Morad Amarzguioui}
\email{morad@astro.uio.no}
\affiliation{Institute of Theoretical Astrophysics, University of
  Oslo, PO Box 1029 Blindern, 0315 Oslo, Norway}

\author{\O{}yvind Gr\o{}n}
\email{oyvind.gron@iu.hio.no}
\affiliation{Oslo College, Faculty of Engineering, Cort Adelersgt.~30,
  0254 Oslo, Norway} 
\affiliation{Department of Physics, University of Oslo, PO Box 1048
  Blindern, 0316 Oslo, Norway}

\begin{abstract}
Recently, there have been suggestions that the apparent accelerated
expansion of the universe is not caused by repulsive gravitation due
to dark energy, but is rather a result of inhomogeneities in the
distribution of matter. In this work, we investigate the behaviour of
a dust dominated inhomogeneous Lema\^itre-Tolman-Bondi universe model,
and confront it with various astrophysical observations. We find that
such a model can easily explain the observed luminosity
distance-redshift relation of supernovae without the need for dark
energy, when the inhomogeneity is in the form of an underdense bubble
centered near the observer. With the additional assumption that the
universe outside the bubble is approximately described by a
homogeneous Einstein-de Sitter model, we find that the position of the
first CMB peak can be made to match the WMAP observations. Whether or
not it is possible to reproduce the entire CMB angular power spectrum
in an inhomogeneous model without dark energy, is still an open
question.
\end{abstract}

\maketitle

\section{Introduction}
The first indications that the universe is presently in a state of
accelerated expansion were given by J.-E. Solheim as far back as in
1966 \cite{solheim66}. Using the observed luminosity of several
cluster galaxies he found that the model giving the best fit to the
data was one with a non-vanishing cosmological constant and negative
deceleration parameter. It is, however, only after the more recent
observations of the luminosity of supernovae of type Ia (SNIa) that
this claim has grown in popularity. The first SNIa observations  
supporting this claim were those of Riess et al. in 1998
\cite{riess98} and Perlmutter et al. in 1999 \cite{perlmutter99}.
Since then, more recent observations of supernovae seem to strengthen
this claim even further \cite{tonry03, knop03, Riess:2004nr}. Other
independent observations that appear to favour the picture of a
universe in a phase of accelerating expansion, are the measurements of
the anisotropies in cosmic microwave background (CMB) temperature
\cite{Spergel:2003cb} and the galaxy surveys \cite{tegmark03}. With
these observations in mind, the current period of accelerated
expansion seems to be well-established. The physical mechanism that
drives this accelerated expansion is, however, still an open question.
It is usually ascribed to an exotic energy component dubbed \emph{dark
  energy}, whose nature remains a mystery.

Recently, there have been several papers discussing the possibility
that the apparent accelerated expansion of the universe is not caused
by this mysterious dark energy, but rather by inhomogeneities in the
distribution of matter. Most of these papers look at the backreaction effects 
arising from perturbing homogeneous models, and try to explain the accelerated 
expansion as corrections to the zeroth order evolution from the higher-order,
inhomogeneous terms (see e.g. \cite{barausse:2005,Kolb:2005me,Wiltshire:2005fw,
 Carter:2005mu, Moffat:2005jc,Buchert:2005kj,Kolb:2005da}). 
However, several papers criticizing some of this work have appeared 
\cite{Flanagan:2005,Geshnizjani:2005,Hirata:2005,Siegel:2005,Ishibashi:2005}.

Another approach is to look at inhomogeneities of a larger scale in the form of underdense bubbles. The basic
idea behind this line of explanation is that we live in an underdense
region of the universe, and the evolution of this underdensity is what
we perceive as an accelerated expansion. An analysis of early
supernova data by Zehavi et al. gave the first indications that there
might indeed exist such an underdense bubble centered near us 
\cite{zehavi98}.

Specific models that give rise to such underdensities have been
studied previously in the form of a local homogeneous void
\cite{Tomita:1999rw,Tomita:2001,Tomita:2000jj}. In these works both
the underdensity and the region outside it are assumed to be
perfectly homogeneous Friedmann-Robertson-Walker (FRW) models with a
singular mass shell separating the two regions. The inhomogeneity
manifests itself as a discontinuous jump at the location of the mass
shell.

In this article, we wish to investigate more realistic models where
there is a continuous transition between the inner underdensity and
the outer regions. Therefore we consider an isotropic but
inhomogeneous dust dominated universe model, where the inhomogeneity
is spherically symmetric. The  model can then be described within the
Lema\^i{}tre-Tolman-Bondi (LTB) class of spherically symmetric universe
models \cite{Lemaitre:1933,Tolman:1934,Bondi:1947}. To make contact
with the ordinary FRW models, we assume that the universe is
homogeneous except for an isotropic inhomogeneity of limited spatial
extension, where the transition between these two regions is
continuous. 

In a homogeneous universe, it is possible to infer the time evolution
of the cosmic expansion from observations along the past light cone,
since the expansion rate is a function of time only. In the
inhomogeneous case, however, the expansion rate varies both with
time and space. Therefore, if the expansion rates inferred from
observations of supernovae are larger for low redshifts than higher
redshifts, this must be attributed to cosmic acceleration in a
homogeneous universe, whereas in our case it can simply be the
consequence of a spatial variation, with the expansion rate being
larger closer to us. As shown in \cite{celerier:99}, this results in
an  expression for the luminosity distance-redshift relation where the
inhomogeneity mimics the role of the cosmological constant in
homogeneous models.

However, the supernova observations are not the only data that support
the claim of an accelerating expansion. As mentioned above, CMB
observations also seem to lend support to this claim. Therefore, in
order for our model to be considered realistic, it should also be able
to explain the observed CMB temperature power spectrum. We will not
attempt to reproduce the whole CMB temperature spectrum for our
inhomogeneous model in this paper. For simplicity, we will limit
ourselves to the location of the first acoustic peak. As we will show
in section~\ref{sec:CMB}, it is possible to obtain a very good match
to both the supernova data and the location of the first acoustic peak
simultaneously. In fact, the match to the supernova data is better
than for the $\Lambda$CDM model.

The observed isotropy of the CMB radiation implies that we must be
located close to the center of the inhomogeneity. According to this
picture, we are positioned at a rather special place in the universe.
On the other hand, this model has the attractive feature that there is
no need for dark energy. Also the model is sufficiently simple so that it
can be solved exactly. It is therefore a good toy model for testing
the ideas of inhomogeneities as a solution to the mystery of the dark
energy.


The structure of this paper is as follows: First, we will present our
model in section~\ref{sec:LTB}, parameterized by two functions
$\alpha(r)$ and $\beta(r)$ related to the distribution of matter and
spatial curvature, respectively. Still in section~\ref{sec:LTB}, we
present the formalism needed in order to obtain the luminosity
distance-redshift relation for spherically symmetric, inhomogeneous
models. In section~\ref{sec:CMB} we discuss the physics behind the
first peak of the CMB spectrum and define a shift parameter that
quantifies the deviation of the location of this peak relative to that
of the concordance $\Lambda$CDM model. In section~\ref{sec:results} we
present the results from the confrontation of our model with the
physical tests presented in the preceding sections, and discuss
briefly the possibility of using the recently detected baryon
oscillations in the matter power spectrum to constrain the model even
further. Finally, in section~\ref{sec:conclusion} we summarize our
work.

\section{Spherically symmetric, inhomogeneous universe models}
\label{sec:LTB} 

The line element for a general, spherically symmetric, inhomogeneous
universe model may be written 
\be
\label{eq:metric}
ds^2 = -dt^2 + X^2(r,t)dr^2+R^2(r,t)d\Omega^2 \, .
\ee

The Einstein equations are
\be
G_{\mu\nu} = R_{\mu\nu}-\frac{1}{2}g_{\mu\nu}R = \kappa T_{\mu\nu} 
\ee
where $\kappa = 8\pi G$ and the energy-momentum tensor is assumed to
be $T_{\mu\nu} = \mbox{diag}(\rho,0,0,0)$, i.e. containing matter
only. 
 
Solving the equation $G_{01} = 0$ gives
\be
X(r,t) = \frac{R'(r,t)}{\sqrt{1+\beta(r)}}
\ee
where $\beta(r)$ is an arbitrary function of $r$. Throughout this
paper, we will use a $' = d/dr$ to denote differentiation with respect
to $r$ and $\dot{ } = d/dt$ for differentiation with respect to $t$. 

The Einstein equations for the dust dominated Lema\^{i}tre-Tolman-Bondi
universe models take the form 
\bean
\label{eq:friedman1}
H_\bot^2 + 2H_\parallel H_\bot - \frac{\beta}{R^2}-\frac{\beta'}{RR'} &=&
\kappa \rho\\ 
\label{eq:friedman2}
-6H_\bot^2q_\bot +2H_\bot^2 -2\frac{\beta}{R^2}-2H_\parallel H_\bot +
\frac{\beta'}{RR'} &=& -\kappa\rho 
\eean
where $H_\bot = \dot{R}/{R}$, $ H_\parallel = \dot{R}'/R'$ and $q_\bot =
-R\ddot{R}/\dot{R}^2$.  

Adding Eqs.~(\ref{eq:friedman1}) and (\ref{eq:friedman2}), we obtain
\be
\label{eq:Rddot}
2R\ddot{R} + \dot{R}^2 = \beta
\ee
Integration of this equation leads to
\be
\label{eq:Hbot}
H_\bot^2 = \frac{\beta}{R^2}+\frac{\alpha}{R^3}
\ee
where $\alpha$ is a function of $r$. 

Hence, the dynamical effects of $\beta$ and $\alpha$ are similar to
those of curvature and dust, respectively.

Differentiating Eq.~(\ref{eq:Hbot}) with respect to $r$ and inserting
the result into Eq.~(\ref{eq:friedman2}), we obtain the density
distribution as 
\be
\kappa \rho = \frac{\alpha'}{R^2 R'}
\ee

Substituting Eqs.~(\ref{eq:Rddot}) and (\ref{eq:Hbot}) into the
expression for the deceleration parameter yields 
\be
q_\bot = \frac{1}{2}\frac{\alpha}{R\dot{R}^2} =
\frac{1}{2}\frac{\alpha}{\alpha+\beta R} 
\ee
Obviously, this quantity is non-negative (since $\alpha \geq 0$) and
equal to the usual value $q_\bot = 0.5$ for a spatially flat, dust
dominated universe. Thus, an inhomogeneous, dust dominated universe
cannot be accelerating in the sense of having a negative deceleration
parameter.

Since we are interested in the late time behaviour of this model, we
will define $t=0$ as the time when photons decoupled from matter, i.e.
the time of last scattering. Furthermore, we define $R(r,t=0) \equiv
R_0(r)$ and introduce a conformal time $\eta$ by $\beta^{1/2}dt =
Rd\eta$. Integrating Eqs.~(\ref{eq:friedman1}) and
(\ref{eq:friedman2}) with $\beta>0$ yields
\bean
R &=& \frac{\alpha}{2\beta}(\cosh \eta - 1)\nonumber\\
  &&+R_0\left[\cosh \eta +
  \sqrt{\frac{\alpha+\beta R_0}{\beta R_0}}\sinh \eta\right]\\ 
\sqrt{\beta}t &=& \frac{\alpha}{2\beta}(\sinh \eta-
\eta)\nonumber\\&&
+R_0\left[\sinh \eta + \sqrt{\frac{\alpha+\beta R_0}{\beta 
R_0}}\left(\cosh \eta   -1\right)\right] 
\eean
which is an {\it exact} solution of Einstein's equations for this
class of models. 

The ordinary dust dominated solution for a universe with negative
spatial curvature is found by choosing 
\be
\alpha = H_0^2 \Omega_{m0} r^3, \quad \beta = H_0^2 \Omega_{k0} r^2
\quad \mbox{and} \quad R_0 = 0 
\ee
yielding
\bean
\label{eq:homogen}
R &=& \frac{1}{2}\frac{\Omega_{m0} r}{\Omega_{k0}}(\cosh \eta -1)
\equiv r a(\eta)\\ 
t &=& \frac{1}{2 H_0}\frac{\Omega_{m0}}{\Omega_{k0}^{3/2}} (\sinh \eta
- \eta) \equiv t(\eta) 
\eean
and $\kappa\rho = 3H_0^2 \Omega_{m0} a^{-3}$. Here, $a(\eta)$ is
recognized as the scale factor in the FRW model, while $\Omega_{m0}$
and $\Omega_{k0}$ are the matter and curvature density today,
respectively. 

Since we are interested in studying a universe with an underdensity at
the center, we choose the $\alpha$ and $\beta$ functions so that they
interpolate between two such homogeneous solutions: 
\bean
\alpha(r) &=& H_{\perp,0}^2r^3\left[ \alpha_0-\Delta \alpha
  \left(\frac{1}{2}-\frac{1}{2}\tanh \frac{r-r_0}{2\Delta r}\right)
  \right]\\  
\beta(r) &=& H_{\perp,0}^2r^2\left[ \beta_0 - \Delta \beta 
  \left(\frac{1}{2}-\frac{1}{2}\tanh \frac{r-r_0}{2\Delta r}\right)
  \right]
\eean
Here, $H_{\perp,0}$ is the value of the transverse Hubble parameter in
the outer homogeneous region today, while $\alpha_0$ and $\beta_0$ are
given by the matter and curvature density in this region,
respectively. Furthermore, $\Delta \alpha$ and $\Delta \beta$ specify 
the differences in the parameters between the two regions, and $r_0$
and $\Delta r$ specify the position and width of the transition.

The function $R_0(r)$ can be chosen freely by a suitable choice of
coordinates $r$ (if the universe has a finite size at $t=0$). To match
our solution to a homogeneous FRW solution in the outer region, we
choose $R_0 = a_* r$, where $a_*$ is the scale factor of the
homogeneous model at recombination.  

To relate the $\alpha$ and $\beta$ functions to observable quantities,
we define relative matter and curvature densities from the generalized
Friedmann equation 
(\ref{eq:friedman1}) as
\bean
\Omega_m &=& \frac{\kappa \rho}{H_\bot^2+2H_\parallel H_\bot}\\
\Omega_k &=& 1-\Omega_m
\eean  
Note that for the homogeneous case, with $H_\bot = H_\parallel$, these
expressions coincide with the usual definitions.  

Furthermore, we need to find the luminosity distance-redshift
relation in this model for comparison with supernova observations. 
The photons arriving at $r=0$ today (defined as $t=t_0$) follow a path
$t = \hat{t}(r)$ given by 
\be
\frac{d\hat{t}}{dr} = -\frac{R'(r,\hat{t})}{\sqrt{1+\beta}}
\ee
with $\hat{t}(0) = t_0$. Following Iguchi et al. \cite{Iguchi:2002},
we find the redshift $z = z(r)$ of these photons from  
\be \label{Eq:zr}
\frac{dz}{dr} = (z+1)\frac{\dot{R}'(r,\hat{t})}{\sqrt{1+\beta}}
\ee
with the initial condition $z(0) = 0$. The position of the last
scattering surface (i.e. the position of the CMB photons that we
observe today, at the time of last scattering) is given by
$\hat{t}(r_*) = 0$, and we define $t_0$ by $z(r_*) = z_* \simeq 1100$.
An accurate formula for $z_*$ in terms of the matter contents of the
universe has been given by Hu and Sugiyama \cite{Hu:1995en}.

The luminosity distance is then given by the usual expression
\be
d_L(z) = (1+z)^2 R(r,\hat{t})
\ee
and the angular diameter distance is
\be
\label{eq:angdist}
d_A(z) = R(r,\hat{t})
\ee

\section{The cosmic microwave background} \label{sec:CMB}

To confront our model with observations of the CMB, we would in
principle need to study perturbations in an inhomogeneous universe.
However, since our model is homogeneous outside a limited region at
the center, we will assume that the evolution of perturbations is
identical to that in a homogeneous universe until the time of last
scattering. This means that we can use the ordinary results for the
scale of the acoustic oscillations at the last scattering surface. On
the other hand, the angular diameter distance, which converts this
scale to a corresponding angle on the sky, {\it is} sensitive to the
inhomogeneity at the center. As a simple test we will use the position
of the first peak in the CMB spectrum to constrain our inhomogeneous
models.


The position of the $m$-th Doppler peak in the CMB spectrum can be
written as \cite{Hu:2000ti}
\be
\label{eq:phim}
l_m = (m-\varphi_m)l_A
\ee
where $l_A$ is the acoustic scale and $\varphi_m$ is a small shift
mainly due to the projection of the three-dimensional temperature
power spectrum onto a two-dimensional angular power spectrum. 

The acoustic scale is given by
\be
l_A = \pi \frac{d_A}{r_s}
\ee
where $d_A$ is the angular diameter distance to the last scattering
surface and $r_s$ is the sound horizon at recombination. In a standard 
FRW cosmology, these two quantities are approximately given by (in
comoving coordinates) 
\be \label{eq:rs}
r_s = \int_0^t \frac{c_s(t')}{a(t')}dt'
\ee
and
\be
d_A = \frac{1}{H_0\sqrt{|\Omega_{k0}|}}S_k\left[H_0\sqrt{|\Omega_{k0}|}\int_0^{z_*}
  \frac{dz}{H(z)}\right] 
\ee
where $c_s$ is the sound speed of the baryon-photon plasma prior to
recombination and $z_* = 1/a_*-1$ is the redshift at the time of
recombination. The function $S_k$ depends on the spatial curvature and
is defined as
\be
S_k(x) = \left\{\begin{array}{lcc}
\sin x & , & \Omega_k < 0\\
x      & , & \Omega_k = 0\\
\sinh x& , & \Omega_k > 0
\end{array}\right.
\ee
A fitting formula for the dependence of $\varphi_1$ on $\omega_b$ and
$\omega_m$, where $\omega_i = \Omega_i h^2$ is the density of energy
component $i$, can be found in Ref.~\cite{Doran:2001yw}. The formula can
be written as 
\be
\varphi_1 = a_1 \left[\frac{\omega_\gamma}{\omega_m a_*}\right]^{a_2}
\ee
where $a_1$ and $a_2$ are given by
\bean
a_1 &=& 0.286+0.626\omega_b\\
a_2 &=& 0.1786-6.308\omega_b+174.9\omega_b^2-1168\omega_b^3
\eean

To reduce the effect of the approximations made in the above formulae,
we will introduce a shift parameter that measures the position of the
first Doppler peak for a given model {\it relative} to the concordance
$\Lambda$CDM model. That is, we define 
\be
\label{eq:shift}
\mathcal{S} = l_1 / l_1^{\Lambda CDM}
\ee
where $l_1^{\Lambda CDM}$ is the peak position for the current
concordance model, with $\omega_k = 0$, $\omega_m = 0.135$, $\omega_b
= 0.0224$, $\omega_\gamma = 4.2\cdot 10^{-5}$ and $\omega_\Lambda =
0.368$. To be in agreement with the WMAP observations, the shift
parameter should be within the range $\mathcal{S} = 1.00 \pm 0.01$. In
fact, the relative error in the peak position from the WMAP data
\cite{Page:2003} is $0.8/220.1 \simeq 4 \cdot 10^{-3}$. However, the
approximations made in the formula in Eq.~(\ref{eq:shift}) are
probably of the same order of magnitude. Therefore it is safe to say
that models with $|\mathcal{S} -1| > 0.01$ are ruled out, whereas
models with $\mathcal{S}$ within a percent of the $\Lambda$CDM value
are probably worth a closer look. After all, there is still a long way
to go from the correct position of one peak to a perfect match with
the entire CMB angular power spectrum.

In addition to the correct peak position, $\omega_b$ should be within
the range predicted by Big Bang nucleosynthesis \cite{Burles:2000zk},
$\omega_b = 0.020 \pm 0.002$. For simplicity, we will use the best-fit
value given by the WMAP team \cite{Spergel:2003cb}, $\omega_b = 0.0224$.   

\begin{figure}
\begin{center}
\includegraphics[width=9.0cm]{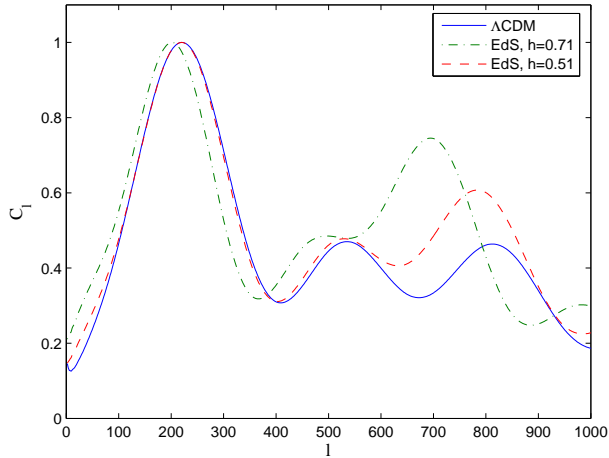}
\end{center}
\caption{CMB angular power spectra for the $\Lambda$CDM model and two
Einstein-de Sitter models, normalized to the height of the first Doppler peak} 
\label{fig:shift}
\end{figure}

Inserting the value for $l_1^{\Lambda CDM}$, Eq.~(\ref{eq:shift}) becomes
\be
\label{eq:shift2}
\mathcal{S} = 0.01419 (1-\varphi_1)\frac{d_A}{r_s}
\ee
As an example, an Einstein-de Sitter (EdS) model with $\Omega_m = 1$, Hubble
parameter $h=0.71$ and $\omega_b = 0.0224$ has $\mathcal{S} = 0.916$,
whereas the same model with $h=0.51$ has $\mathcal{S} = 0.998$. In
comparison, CMBFAST \cite{seljak:1996} yields the values $\mathcal{S} =
0.914$ and $\mathcal{S} = 0.998$ for these two models. As we can see,
the formalism accurately describes the position of the first peak. The
CMB temperature power spectra of these three models are plotted in
figure~\ref{fig:shift}. Note that the second and third peaks will also have 
approximately correct positions when the shift parameter is close to 1, since
the $\varphi_m$'s in Eq.~(\ref{eq:phim}) are relatively small. For instance, 
the EdS model with $h=0.51$ has $l_2 = 530$ and $l_3 = 784$, whereas the 
best-fit values from \cite{Jones:2005} are $l_2 = 529$ and $l_3 = 781$. Also note that 
the relative error in the location of the second and third peak
are larger than for the first peak, at around 3\%. 

The relative heights of
the peaks, on the other hand, have a more complicated dependence on the 
parameters of the model, see e.g. \cite{Hu:2000ti}. We will postpone discussing 
these features of the CMB spectrum until we have a better understanding 
of the evolution of perturbations in an inhomogeneous model. Note, however, that it is possible to make 
matter-dominated homogeneous models that fit the observed CMB spectrum, see e.g. \cite{Blanchard:2003}. 

In our case, we must use expression~(\ref{eq:angdist}) for the angular diameter
distance to the last scattering surface when we calculate the shift parameter
in Eq.~(\ref{eq:shift2}). On the other hand, we can still use the expression
for the sound horizon as defined in the homogeneous case in Eq.~(\ref{eq:rs}),
since our model is assumed to be homogeneous close to the last scattering
surface.

\section{Results} \label{sec:results}

When going from a homogeneous to an inhomogeneous universe model, the
parameters describing the model ($\omega_m$ and $\omega_k$) become
functions of $r$. This means that we introduce, in principle, an
infinite number of new degrees of freedom. However, for the purpose of
studying the possibility of explaining the current observations
without introducing dark energy into the model, we have restricted
ourselves to a very simple ``toy model'': An underdense region close
to us, surrounded by a flat, matter dominated universe. This means
that we must choose $\alpha_0 = 1$ and $\beta_0 = 0$. Furthermore, we
put $\Delta \alpha = -\Delta \beta$. This leaves four parameters,
$\Delta \alpha$, $r_0$, $\Delta r$ and the physical Hubble parameter
at the origin, $H_\parallel(0,t_0) = \unit[100 h]{\frac{km}{s\cdot Mpc}}$, to
be fitted to the observations.

Let us first focus on the two main observations: The supernova Hubble
diagram and CMB angular power spectrum. A good fit to the supernova
data requires the Hubble parameter inside the underdensity, $h_{\text{in}}$
to be around $h_{\text{in}} \simeq 0.65$. On the other hand, a good fit to
the CMB spectrum for a flat matter dominated model requires the Hubble
parameter outside the underdensity to be $h_{\text{out}} \simeq 0.5$. This
more or less determines the two parameters $\Delta \alpha$ and $h$.

\begin{figure}
\begin{center}
\includegraphics[width=9.0cm]{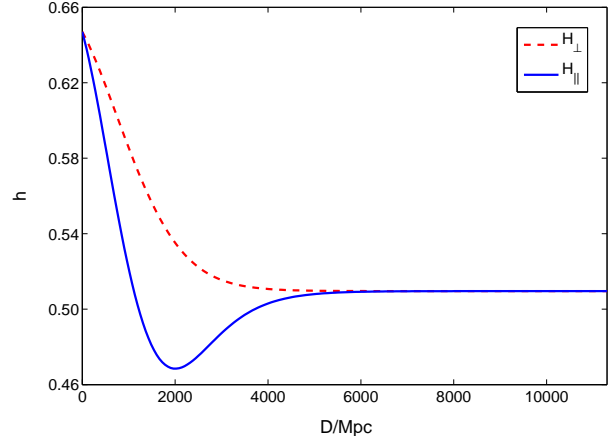}
\end{center}
\caption{The spatial variation of the Hubble-parameters at $t=t_0$}
\label{fig:hubble}
\end{figure}

\begin{figure}
\begin{center}
\includegraphics[width=9.0cm]{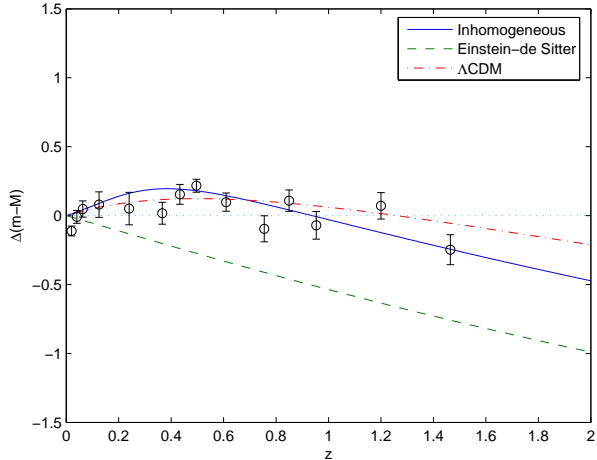}
\end{center}
\caption{Distance modulus vs. redshift for our standard
  model together with supernova observations}
\label{fig:sn1a}
\end{figure}

Next, the shape of the transition between the underdense and the
homogeneous region is specified by $r_0$ and $\Delta r$. These values
dictate the redshift-magnitude relationship, and must be chosen to fit
the supernova Hubble diagram. There are lots of choices for the
parameters that give a very good fit to both the supernovae and the
position of the first acoustic peak in the temperature power spectrum.
However, we want the underdensity in our model to be such that the
matter density is compatible with the current model independent
observations of $\Omega_{m0}$. An excellent candidate for such
observations is the mass-to-light ratio measurements made by the 2dF
team \cite{Cross:2000ut}. These yield $\Omega_{m0} = 0.24 \pm 0.05$ from
observations of galaxies with redshifts $z < 0.12$. We will therefore
choose the free parameters such that the mass density parameter at the
origin is within this range in addition to giving a good fit to the
supernova measurements and the CMB peak. The model which we adopt
as our ``standard model'' gives a matter density at the center of the
underdensity of $\Omega_{m0} = 0.20$. A plot of the spatial variation
today of the Hubble parameters of our standard model is given in
figure~\ref{fig:hubble}. Furthermore, a plot of the distance modulus of
this model together with the supernova observations can found in
figure~\ref{fig:sn1a} . 

Note that the $\chi^2$-value for our model is $\chi^2 = 176.5$, when
compared to the ``gold'' dataset of Riess {\it et al.}
\cite{Riess:2004nr}. This is slightly better than that of the concordance
$\Lambda$CDM model \cite{Riess:2004nr}, $\chi^2_{\Lambda CDM} = 178$.

\begin{figure}
\begin{center}
\includegraphics[width=9.0cm]{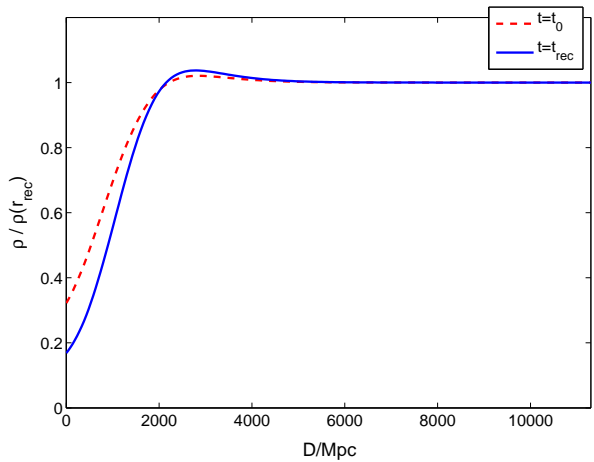}
\end{center}
\caption{The evolution of the underdensity in our standard model}
\label{fig:omegam}
\end{figure}

The spatial shapes of the underdensity at the initial time and today
are plotted in figure~\ref{fig:omegam} as functions of the physical
distance today. This illustrates the time evolution of the underdensity.
As we can see, the shape stays almost constant. This is due to the
Hubble parameters $H_\parallel$ and $H_{\perp}$ being roughly constant in space.
%

Although the matter distribution is clearly inhomogeneous close to the observer,
we wish to point out that this does not necessarily contradict the data from the
galaxy surveys. One often hears the claim that these surveys show the local distribution 
of matter to be homogeneous. However, it is probably more correct to say that they are shown
to be compatible with a homogeneous universe rather than actually proving it. The key
point here is that in order to determine for example the number counts of galaxy clusters, 
one needs to make an assumption about galaxy evolution and how likely it is to observe a 
galaxy with a certain luminosity at a certain redshift. As pointed out
in e.g. \cite{mustapha:1998}, one usually assumes a homogeneous
universe in order to deduce the effects of source evolution. Therefore, using this deduced evolution to claim observed homogeneity in the number counts
amounts only to circular argumentation. Furthermore, it is explicitly
shown in \cite{mustapha:1998} that 
given any LTB model it is always possible to find a source evolution that agrees with the
observed number counts. 

The inhomogeneity at the center gives only a minor change of the
angular diameter distance to the last scattering surface. In fact, our
model has $d_A = \unit[10.4]{Mpc}$, which is the same value as we find
for the Einstein-de Sitter model with $h = 0.51$. (Note that these
values are physical, not comoving distances). Our model,  with Hubble
parameter $h \simeq 0.51$ in the homogeneous region, thus yields a CMB
angular power spectrum very similar to the one plotted in 
figure~\ref{fig:shift}, at least for large $l$ values. Using the
formula~(\ref{eq:shift2}), we find $\mathcal{S} = 1.006$, i.e. an
almost perfect match for the position of the first Doppler peak. For
smaller $l$ values, the CMB pattern will be affected by our position
relative to the center of the underdensity. This has been studied in
the previously mentioned void model of Tomita \cite{Tomita:1999rw}, 
who concluded that relatively large displacements from the center of
the underdensity were fully consistent with the observed CMB dipole
and quadrupole.
Furthermore, J. Moffat \cite{Moffat:2005jc} argues that such a
displacement could even explain the detected alignment of the CMB
quadru- and octopole \cite{Schwarz:2004}.

A rough estimate of the apparent peculiar velocity for an off-center observer is \cite{Tomita:1999rw}
\be
v_p \simeq \left(h_{\parallel,\text{in}}-h_{\parallel,\text{out}}\right)l_0\cdot \unit[100]{km/s}
\ee
where $l_0$ is the distance from the observer to the center, measured in Mpc.
If we for instance require  that $v_p$ must be less than the estimated peculiar 
velocity of the local group \cite{kogut:1993}, which is of the order of \unit[600]{km/s}, 
this means that the observer must be within $\unit[40]{Mpc}$ from the center of the inhomogeneity. Even stronger constraints might be obtained by
considering the peculiar velocities of nearby clusters, see e.g. \cite{hudson:2004}.

Recently, Eisenstein et al. announced the detection of baryon
oscillations in the SDSS galaxy power spectrum 
\cite{Eisenstein:2005su}. This represents additional, independent data
that can be used to constrain our model even further. The physical
length scale associated with these oscillations is set by the sound
horizon at recombination. Measuring how large this length scale
appears at some redshift in the galaxy power spectrum allows us to
constrain the time evolution of the universe from recombination to
the time corresponding to this redshift
\cite{Hu:1995en,Eisenstein:1997ik,Hu:2003ti,Blake:2003rh,Linder:2003ec,Linder:2005tg}.

A length scale quoted by Eisenstein et al. is the ratio of the
effective distance to the chosen redshift in the galaxy survey to the
angular diameter distance to the last scattering surface,
\begin{equation}
  R_{0.35}=\frac{d_V(z_{sdss})}{d_A(z_*)}\,,
\end{equation} 
where $z_{sdss}=0.35$. The effective distance $d_V$ is defined in
Eq.~(2) in Ref.~\cite{Eisenstein:2005su} as a mix of radial and
angular distance, to take into account that these scale differently.
The value they measure for this ratio is $R_{0.35}=79.0\pm2.9$. Note
that this value differs from that quoted in \cite{Eisenstein:2005su}.
The reason for this is that we've chosen to give the distances $d_V$
and $d_A$ in physical coordinates, while Eisenstein et al. quote them
as comoving.

Calculating this ratio for our model we find the value
$R_{0.35}^{inhom}=107.1$. Comparing this value with that quoted by
Eisenstein et al., one might be tempted to claim that the model is
ruled out. However, in order to say something conclusive using this
constraint, we need to be sure that the ``measured'' value of
$R_{0.35}$ is model independent. But when the authors derived this
constraint they assumed a $\Lambda$CDM model. This makes it a little
unclear how to use this constraint for non-$\Lambda$CDM models, or,
indeed, whether it is even possible to use it for such models.
Ideally, one would need to repeat the analysis of Eisenstein et al.
assuming our inhomogeneous model as base model. We will therefore be
careful not to rule out the model based on this parameter alone. 

The main features of our standard model are summarized in
table~\ref{tab:model}.
\squeezetable
\begin{table}[t]
\begin{tabular}{lcc}
\hline 
Description & Symbol & Value\\
\hline 
Density contrast parameter & $\Delta \alpha$ & 0.90 \\
Transition point & $r_0$ &  1.35 Gpc \\
Transition width & $\Delta r/r_0$ & 0.40 \\
Fit to supernovae & $\chi^2_{SN}$ & 176.5\\
Position of first CMB peak & $\mathcal{S}$ & 1.006 \\
Age of the universe & $t_0$ & $\unit[12.8]{Gyr}$\\
Relative density inside underdensity & $\Omega_{m,in}$ & 0.20 \\
Relative density outside underdensity & $\Omega_{m,out}$ & 1.00 \\
Hubble parameter inside underdensity & $h_{in}$ & 0.65 \\
Hubble parameter outside underdensity & $h_{out}$ & 0.51 \\
Physical distance to last scattering surface & $D_{LSS}$ & 11.3 Gpc\\
Length scale of baryon oscillation from SDSS & $R_{0.35}$ & 107.1 \\
\hline
\end{tabular}
\caption{The parameters and features of our adopted standard
  inhomogeneous model}
\label{tab:model}
\end{table}
Note that the age of the universe is $\unit[12.8]{Gyr}$ in our model. This is significantly less than the value for the concordance $\Lambda$CDM model, $\unit[13.7]{Gyr}$, but it
is still in agreement with observations of globular clusters \cite{Krauss:2003}, which put a lower limit of $\unit[11.2]{Gyr}$ on the age of the universe.








\section{Summary and conclusions} \label{sec:conclusion}

The main goal of this paper has been to present a simple model with the
ability to explain the apparent accelerated expansion of the universe
without the need to introduce dark energy. Inspired by the recent
discussions about the possibility of explaining the apparent acceleration
by inhomogeneities in the matter distribution, we have studied a model
where the observer is assumed to be situated near the center of an
underdense bubble in a flat, matter dominated universe. If this model
is realistic, we live in a perturbed Einstein-de Sitter universe within 
130 million light years
from the center of an underdensity that extends about 5 billion light
years outwards. Under the assumption of spherical symmetry this
universe is described by the Lema\^itre-Tolman-Bondi space-time. 

The two main observations we sought to explain were the luminosity
distance-redshift relation inferred from SNIa observations and the CMB
temperature power spectrum. These two sets of observations are made at
opposite ends of the redshift spectrum, respectively low redshifts for
the supernovae and high redshifts for the CMB. The fact that our model
is inhomogeneous allows us therefore to choose the geometry and
matter distribution such that the physical conditions are favourable
for explaining the SNIa at low redshift while they at the same time are
favourable for explaining the CMB at high redshifts. We find that a
very good fit to the supernova data is obtained if we allow the
transverse Hubble parameter to decrease with the distance from the
observer. On the other hand, we get a good fit to the location of
the first peak of the CMB power spectrum if we assume the universe
to be flat with a value of $0.51$ for the Hubble parameter outside the
inhomogeneity. Interpolating between these two limiting behaviours we
get a good fit to both the supernova data and the location of the first
peak.

Our model yields a better fit to the Riess data set of supernovae than
the concordance $\Lambda$CDM model. However, for the CMB fit we tested
only for the location of the first peak. Although the model yields a
good fit to this, it does not necessarily mean that it matches the whole
CMB spectrum. Indeed, since the physics responsible for the acoustic
peaks is determined by the pre-recombination era, we would expect the
peaks to look more or less the same as for a flat, homogeneous model
with $h=0.51$. This suggests that our model might fail to explain the
third peak. Furthermore, the model does not appear to be able to explain
the observed length scale of the baryon oscillations in the SDSS matter
power spectrum either, although one may question whether the data
quoted by the SDSS team can be used directly to test our model.

The most powerful way to rule out inhomogeneous universe models would
be to do a full analysis of the evolution of perturbations in these
models. In that way, one could confront the model with both the full
CMB angular power spectrum and the matter power spectrum. Only after
such an analysis is carried out can one say whether our model is ruled
out or if it is a viable alternative to dark energy.

\begin{acknowledgments}
The authors would like to thank M.J. Hudson for helpful discussions.
MA acknowledges support from the Norwegian Research Council through
the project ``Shedding Light on Dark Energy'', grant 159637/V30.
\end{acknowledgments}

\end{document}